\documentclass[%
preprint,
 amsmath,amssymb,
 aps,
]{revtex4-2}

\UseRawInputEncoding
\usepackage{graphicx}
\usepackage{dcolumn}
\usepackage{bm}

\begin{document}

\preprint{}

\title{Passive Synchronization of Nonlocal Franson Interferometry for Fiber-Based Quantum Networks Using Co-propagating Classical Clock Signals}
\author{Xiao Xiang$^{1,2,3}$}
\author{Runai Quan$^{1,2,3}$}
\author{Yuting Liu$^{1,2}$}
\author{Huibo Hong$^{1,2}$}
\author{Bingke Shi$^{1,2}$}
\author{Zhiguang Xia$^{1,2}$}
\author{Xinghua Li$^{1,2}$}
\author{Tao Liu$^{1,2,3}$}
\author{Shougang Zhang$^{1,2,3}$}
\author{Ruifang Dong$^{1,2,3,}$}
\email{dongruifang@ntsc.ac.cn}

\affiliation{$^1$Key Laboratory of Time Reference and Applications, National Time Service Center, Chinese Academy of Sciences, Xi'an 710600, China}
\affiliation{$^2$School of Astronomy and Space Science, University of Chinese Academy of Sciences, Beijing 100049, China}
\affiliation{$^3$ Quantum Precision Measurement Department, Hefei National Laboratory, Hefei 230026, China}





\date{\today}

\begin{abstract}
We demonstrate a robust, high-visibility nonlocal Franson interferometry for fiber-based quantum networks by co-propagating a classical Radio-over-Fiber clock signal with energy-time entangled photon pairs in the same fiber. Utilizing cross-band allocation (O-band for classical, L-band for quantum signals), the spontaneous Raman scattering noise photons are effectively suppressed. At the same time, their environmental delay fluctuations remain highly correlated for common-mode noise cancellation, achieving a passive synchronization with picoseconds precision. Over 50 km of single-mode fiber, this co-propagation enables nonlocal quantum interference with a visibility of \((88.35\pm3.62)\%\), without relying on external dedicated timing infrastructure. This work provides a practical, scalable synchronization solution for metropolitan-scale entanglement-based quantum networks.
\end{abstract}

\maketitle

\section{\label{sec:level1}Introduction}
Quantum entanglement distribution\cite{yin_satellite-based_2017,neumann_continuous_2022,haldar_towards_2023} stands as a foundational pillar for the realization of wide-area quantum networks, which promise revolutionary applications in secure quantum key distribution (QKD) and distributed quantum computing and sensing. Among various photonic degrees of freedom, energy-time entanglement has garnered significant attention for fiber-based metropolitan-scale networks\cite{cuevas_long-distance_2013,xavier_energy-time_2025}, as it benefits from the inherent insensitivity to ubiquitous polarization perturbations and fluctuations in long-distance installed optical fibers. This property translates to superior robustness for entanglement distribution, obviating the need for complex and costly active polarization stabilization systems required for distributing polarization-encoded qubits. This resilience has been utilized in numerous applications, such as quantum time synchronization\cite{Shi_250km_2024,hong_quantum_2024,hong_cascaded_2026} and multiple QKD\cite{liu_high-dimensional_2024,Fan2025} protocols over long-haul fibers.

On the other hand, high-fidelity distribution of energy-time entanglement between distant nodes relies on non-local interference measurement, i.e., the Franson interference \cite{Franson1989, xia_dispersion_2023, fallon_franson_2025}. Such a nonlocal Franson interferometer demands precise time synchronization to align the detection events of entangled photons at separate stations \cite{liu_high-dimensional_2024}. High-precision synchronization protocols, such as the White Rabbit precision time protocol (WR-PTP) or pulsed-laser-based synchronization, can achieve picosecond-level timing accuracy over optical fibers \cite{burenkov_synchronization_2023, mckenzie_clock_2024}. However, these synchronization signals are typically delivered via a dedicated fiber channel, which consumes additional fiber resources. In contrast, co-propagation of classical and quantum signals in a shared fiber offers a resource-efficient path toward integrated quantum network functionalities \cite{Geng_2021,ramesh_HOM_2025}. Nevertheless, experimental demonstrations of classical synchronization signals co-propagating with energy–time entangled photons remain limited, with only a few studies having validated their coexistence with fiber-optic communication \cite{Fan2023, Luo2025}. Moreover, a quantitative framework that defines the benchmark for maintaining high-visibility Franson interference is still lacking.

In this work, we address this gap by introducing and analyzing a robust classical-quantum co-propagation scheme, where a classical Radio-over-Fiber (RoF) signal, carrying a 100 MHz clock signal, is transmitted alongside the energy-time entangled photons through the same optical fiber. To mitigate spontaneous Raman scattering (SpRS) noise from the strong classical signal, we implement a cross-band strategy in which the classical RoF signal is allocated to the O-band (1310 nm), while the quantum signals operate in the L-band (1575 nm). Despite a wavelength separation of 265 nm, the classical and quantum signals exhibit highly correlated propagation behavior under environmental variations along the shared fiber. Benefiting from the common-mode time delay cancellation, we demonstrate nonlocal Franson interference with visibility of \((88.35\pm3.62)\%\) after co-propagation over 50 km of fiber. This performance not only surpasses the typical visibility threshold required for QKD protocols but also validates the effectiveness of the passive synchronization mechanism. By eliminating dependence on external timing infrastructures such as global navigation satellite system (GNSS) or WR-PTP, the proposed scheme offers a cost-effective, fiber-integrated, and readily scalable synchronization solution. Our work thus provides both a practical synchronization framework and a clear performance benchmark, paving a feasible pathway toward real-world deployment of energy-time entanglement-based metropolitan quantum networks.

\section{THEORETICAL DIAGRAM}
\begin{figure}[h]
    \includegraphics[width=8cm]{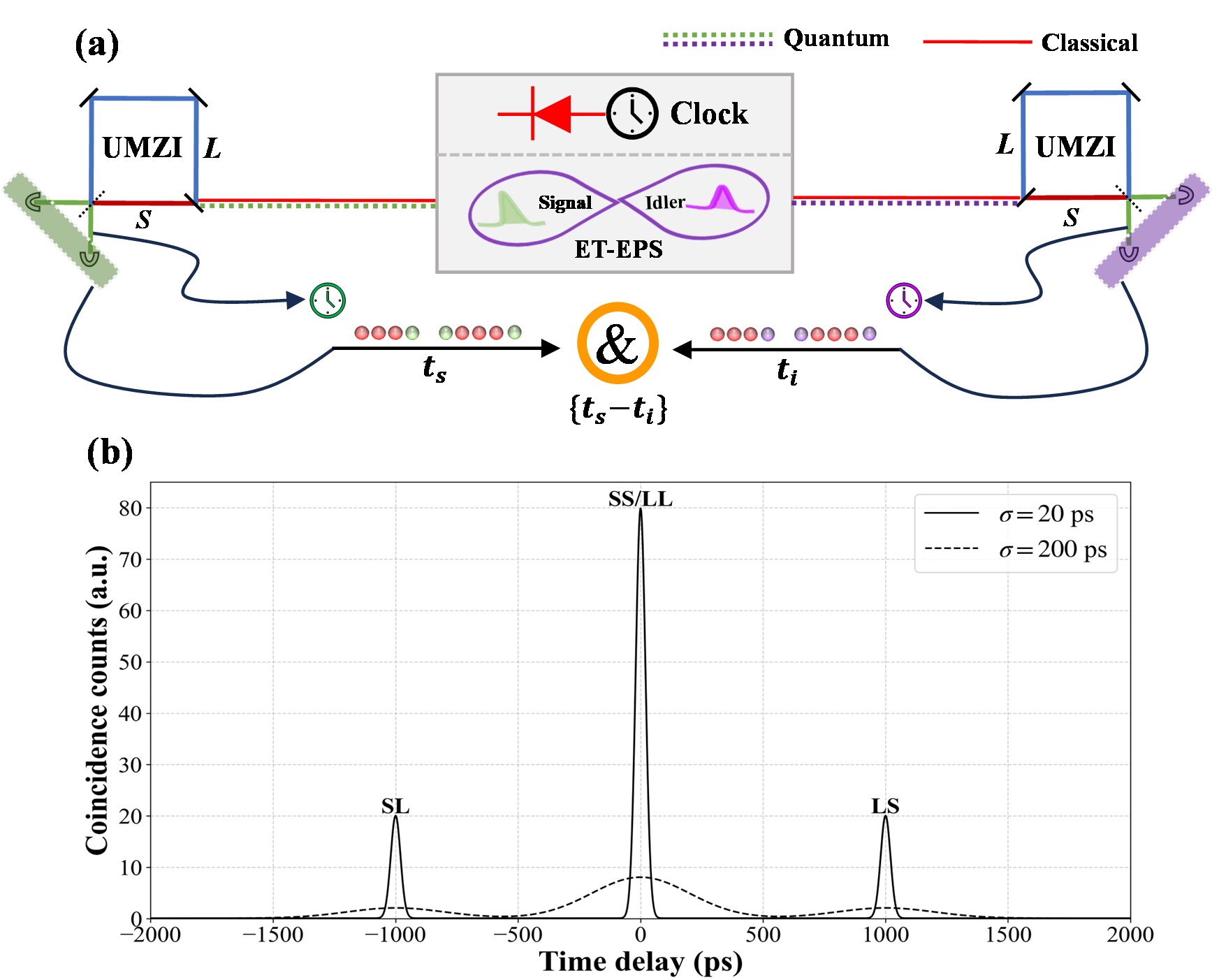}
    \caption{\label{fig:F1} (a) Schematic of a typical nonlocal Franson interferometry with co-propagating classical clock signals. (b) Corresponding coincidence envelopes in the time domain, showing the central interference peak and the two non-interference side peaks.}
\end{figure}

\subsection{Influence of Time Synchronization Error on Franson Interference}
As illustrated in Fig.~\ref{fig:F1}(a), a typical Franson interference setup employs two unbalanced Mach-Zehnder interferometers (UMZIs) to characterize an energy-time entangled photon source (ET-EPS). The signal (s) and idler (i) photons are directed into separate UMZIs, yielding three coincidence envolopes in the time domain, as shown in Fig.~\ref{fig:F1}(b). The left and right peaks are non-interference peaks, corresponding to the distinguishable short-long (SL) and long-short (LS) path combinations, respectively. The central peak results from two-photon interference between the indistinguishable short-short (SS) and long-long (LL) terms. To observe clean two‑photon interference while suppressing single‑photon interference, the difference in path‑length  of each UMZI must satisfy two conditions: it must be longer than the coherence length of the individual signal and idler photons, yet shorter than the coherence length of the pump laser for generating ET-EPS. 

Under ideal conditions without time synchronization errors between the two spatially separated UMZIs, the coincidence peaks are narrow, represented with solid line in Fig.~\ref{fig:F1}(b). It is primarily determined by the timing jitter of the single-photon detection and counting modules, with a standard deviation ($\sigma_0$) typically on the order of tens of picoseconds. In the presence of time synchronization errors, i.e., $\sigma_{\text{sync}}\neq0$, the coincidence peaks ($\sigma_m$) broaden according to:
\begin{equation}
\label{eq:sigma_m}
\sigma_m = \sqrt{\sigma_0^2 + \sigma_{\text{sync}}^2}.
\end{equation}
When the synchronization error is significant, substantial broadening in the coincidence distribution occurs, as illustrated with the dashed line in Fig.~\ref{fig:F1}(b). This can lead to overlap between the central interference peak and the side non-interference peaks, making it hard to select an appropriate coincidence window for achieving high interference visibility.

For subsequent calculations, we define the brightness of the entanglement source as $B$ (in counts per second, cps). Considering the link transmission loss and single-photon detection efficiency, the total transmission efficiencies for the signal and idler photons are $\eta_s$ and $\eta_i$, respectively. The total coincidence count rate is then calculated as $B \eta_s \eta_i$, and the effective coincidence count rate $\text{CC}_e$ within the selected coincidence window ($\tau$) can be expressed using the error function\cite{Neumann2021}:
\begin{equation}
\label{eq:CC}
\text{CC}_e = B \eta_s \eta_i \times \text{erf}\left[\sqrt{\ln(2)} \frac{\tau}{\sigma_m}\right].
\end{equation}
The accidental coincidence count rate $\text{ACC}$ within the same coincidence window is given by:
\begin{equation}
\label{eq:ACC}
\text{ACC} = B^2 \eta_s \eta_i \tau.
\end{equation}
The interference visibility $V$ is then determined by the effective and accidental coincidences:
\begin{equation}
\label{eq:visibility}
V = \frac{(\text{CC}_e + \text{ACC}) - \text{ACC}}{(\text{CC}_e + \text{ACC}) + \text{ACC}} = \frac{\text{CAR}}{2 + \text{CAR}},
\end{equation}
where $\text{CAR}$ stands for the coincidence-to-accidental ratio, 
\begin{equation}
\label{eq:CAR}
\text{CAR} = \frac{\text{CC}_e}{\text{ACC}} = \frac{\text{erf}\left[\sqrt{\ln(2)} \frac{\tau}{\sigma_m}\right]}{B\tau}.
\end{equation}

\begin{figure}
    \includegraphics[width=8cm]{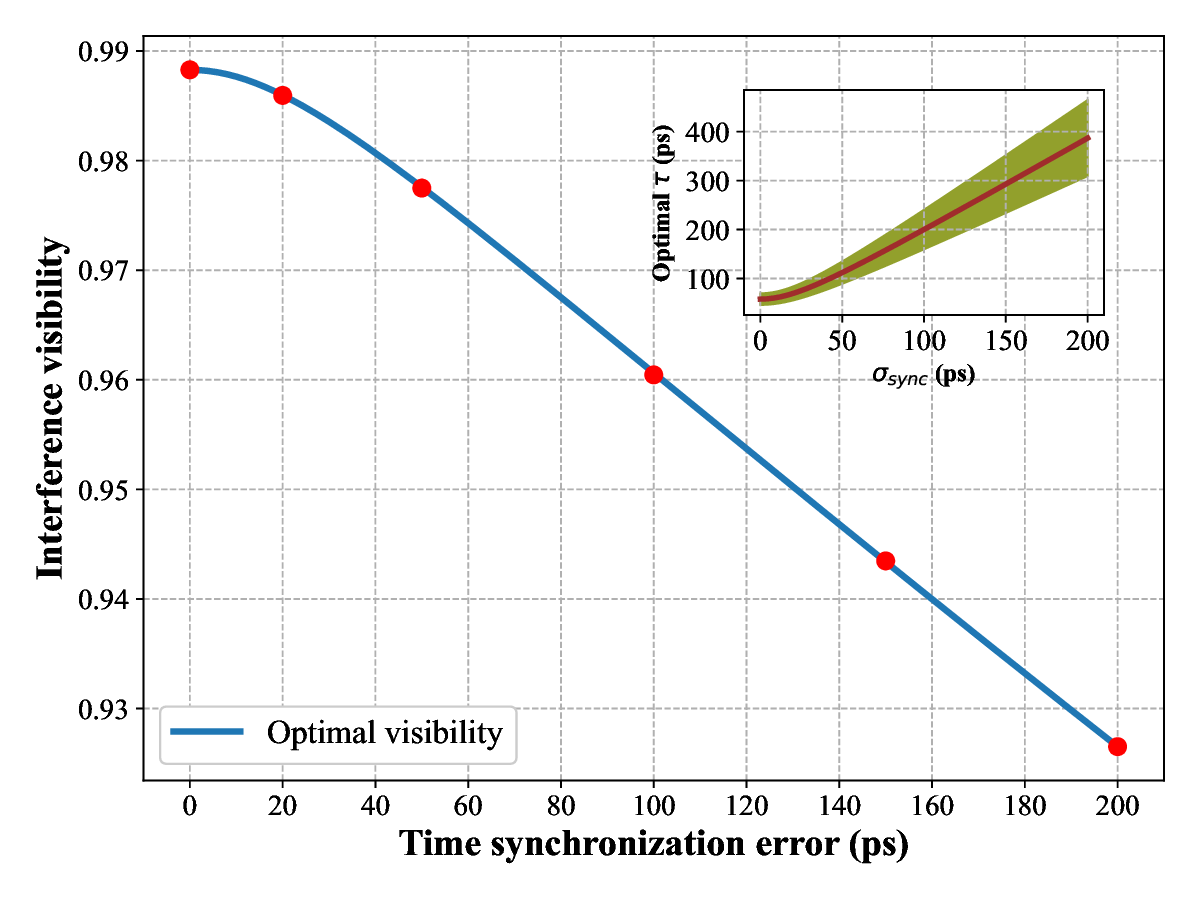}
    \caption{\label{fig:F2} Maximum achievable Franson interference visibility as a function of time synchronization error with $B = 10^8$~cps and $\sigma_0 = 30$~ps. The inset shows the corresponding optimal coincidence window $\tau_{\text{opt}}$ for each $\sigma_{\text{sync}}$. The shaded region in the inset indicates a $\pm 20\%$ tolerance range for coincidence window selection in practical implementations.}
\end{figure}

To investigate the influence of time synchronization error on interference visibility, numerical simulations were performed. Using an entanglement photon source brightness of $B = 10^8$~cps and a detector timing jitter of $\sigma_0 = 30$~ps, we calculated the maximum achievable visibility $V$ as a function of $\sigma_{\text{sync}}$ ranging from 0 to 200~ps, as shown in Fig.~\ref{fig:F2}. The results reveal that interference visibility degrades with increasing $\sigma_{\text{sync}}$, decreasing from approximately 98.8\% at $\sigma_{\text{sync}} = 0$~ps to 92.7\% at $\sigma_{\text{sync}} = 200$~ps. This degradation occurs because time synchronization errors broaden the coincidence distributions, necessitating larger coincidence windows to capture the effective coincidence counts while simultaneously increasing accidental coincidences. The inset of Fig.~\ref{fig:F2} shows the corresponding optimal coincidence window $\tau_{\text{opt}}$ that maximizes $V$ for each $\sigma_{\text{sync}}$. The optimal window grows approximately proportionally to the total timing uncertainty, with $\tau_{\text{opt}}/\sigma_m \approx 2$, providing a practical guideline for adjusting the coincidence window based on measured synchronization performance.

\subsection{Franson Interference with Co-propagating Classical Clock Signal}
Considering a practical scenario where the classical clock signal and the quantum signal (energy-time entangled photons) co-propagate in a single-mode fiber (SMF), the induced SpRS noise power to the quantum channel can be expressed as\cite{burenkov_synchronization_2023}:
\begin{equation}
\label{eq:SpRS}
P(L) = \frac{\left(e^{-\alpha_{\text{c}} L} - e^{-\alpha_{\text{q}} L}\right) \beta_{\lambda_{\text{q}}, \lambda_{\text{c}}} \Delta\lambda_{\text{c}} P_{\text{in}}}{\alpha_{\text{q}} - \alpha_{\text{c}}},
\end{equation}
where \( L \) is the fiber length, \( \alpha_{\text{q}} \) and \( \alpha_{\text{c}} \) are the attenuation coefficients at the quantum and classical signal wavelengths, respectively. The parameter \( \beta_{\lambda_{\text{q}},\lambda_{\text{c}}} \) is the noise conversion coefficient, which characterizes the efficiency of SpRS noise generation and depends strongly on the wavelength allocation. Generally, a larger separation between the classical  wavelength ($\lambda_{\text{c}}$) and the quantum wavelength ($\lambda_{\text{q}}$)  results in a smaller \( \beta \) value. Therefore, in our subsequent simulation and experimental demonstration, the classical clock wavelength is set to \( \lambda_{\text{c}} = 1310\ \text{nm} \), a common telecom window with a moderate fiber attenuation and zero dispersion.
\begin{figure}
    \includegraphics[width=8cm]{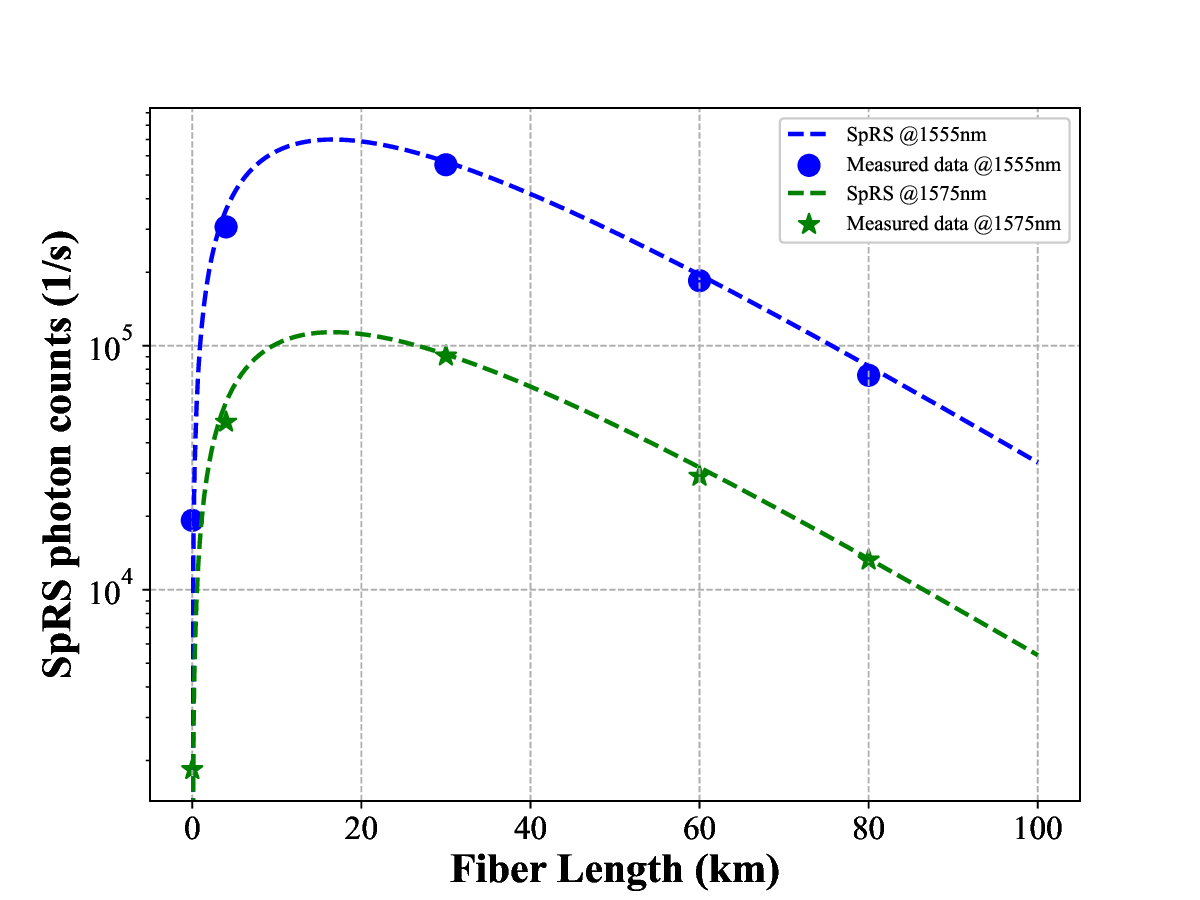}
    \caption{\label{fig:F3} Measured SpRS photon counts as a function of fiber length at 1555~nm (blue dots) and 1575~nm (green stars). The dashed lines represent fits based on Eqn.~(\ref{eq:SpRS}).}
\end{figure}

With an input classical optical power of \( 100\ \mu\text{W} \) (corresponding to a photon flux of \( P_{\text{in}} = 6.58 \times 10^{14}\ \text{photons/s} \)), the SpRS noise photon counts within a \( 1\ \text{nm} \) bandwidth (\( \Delta\lambda_{\text{c}} \)) around the entangled photon wavelengths is measured. 
By analyzing the two characteristic wavelengths of \( 1555\ \text{nm} \) and \( 1575\ \text{nm} \) in Fig.~\ref{fig:F3}, It clearly shows that the SpRS noise photon count decreases significantly with increasing wavelength, particularly from the C‑band to the L‑band. Fitting the data using Eq.~\eqref{eq:SpRS} yields \( \beta_{1555,1310} = 0.4 \times 10^{-23}\ (\text{km}\cdot\text{Hz})^{-1} \) and \( \beta_{1575,1310} = 0.065 \times 10^{-23}\ (\text{km}\cdot\text{Hz})^{-1} \). In addition, the SpRS noise accumulates rapidly with fiber length, reaching a maximum at around \( 20\ \text{km} \). Beyond this point, the noise decreases because fiber attenuation begins to dominate over the nonlinear generation process\cite{burenkov_synchronization_2023}.

Under these SpRS noise conditions and including the detector's dark count rate (DCR), the accidental coincidence count rate is recalculated as:
\begin{equation}
\label{eq:ACC_full}
\text{ACC}_m = (B\eta_s + P(L_s) + \text{DCR}_s)(B\eta_i + P(L_i) + \text{DCR}_i)\tau,
\end{equation}
where \(L_s\) and \(L_i\) represent the fiber propagation distances of the signal and idler photons, respectively. The CAR is modified to
\begin{equation}
\label{eq:CAR_full}
\text{CAR}_m = \frac{B\eta_s\eta_i \times \text{erf}\left[\sqrt{\ln(2)} \frac{\tau}{\sigma_m}\right]}{(B\eta_s + P(L_s) + \text{DCR}_s)(B\eta_i + P(L_i) + \text{DCR}_i)\tau}.
\end{equation}

\begin{figure}
    \includegraphics[width=8cm]{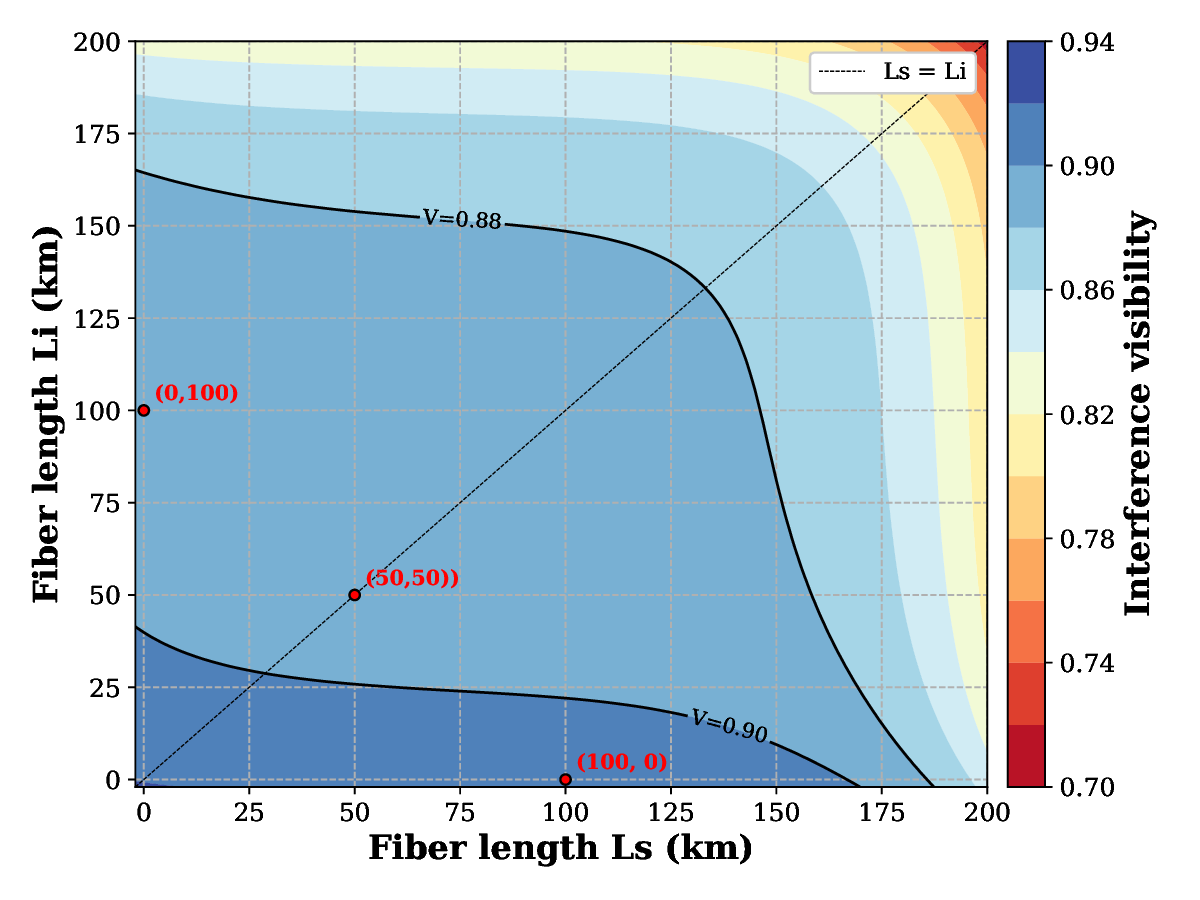}
    \caption{\label{fig:F4} Interference visibility as a function of fiber length for a fixed source brightness $B = 10^8$~cps, single-photon detector DCR = 100 Hz and time synchronization errors of $\sigma_{\text{sync}} = 100$~ps.}
\end{figure}

Based on Eq.~(\ref{eq:CAR_full}), we further analyze the variation of the Franson interference visibility with the fiber transmission distance, taking into account a moderate synchronization error of $\sigma_{\text{sync}} = 100$~ps. In the simulation results presented in Fig.~\ref{fig:F4}, the source brightness is fixed at $B = 10^8$~cps, and the dark count rates of both single-photon detectors are set to $\text{DCR} = 100$~Hz. To illustrate the influence of the quantum photon wavelength allocation on the transmission performance, we set the signal photon wavelength to 1575 nm and the idler photon wavelength to 1555 nm. As clearly shown in Fig.~\ref{fig:F4}, for a fixed total transmission distance, letting the L-band photon travel a longer fiber segment helps achieve higher interference visibility (indicated by red dots), which is attributed to the introduction of less SpRS noise. However, this advantage diminishes as the fiber length increases, because the effective entangled photon flux after fiber attenuation becomes comparable to the detector's DCRs. Therefore, reducing the DCR is essential for extending the achievable transmission distance, as evidenced by recent long-haul fiber experiments \cite{liu_1002km}.

\section{EXPERIMENTAL SETUP}
\begin{figure}
    \includegraphics[width=8.5cm]{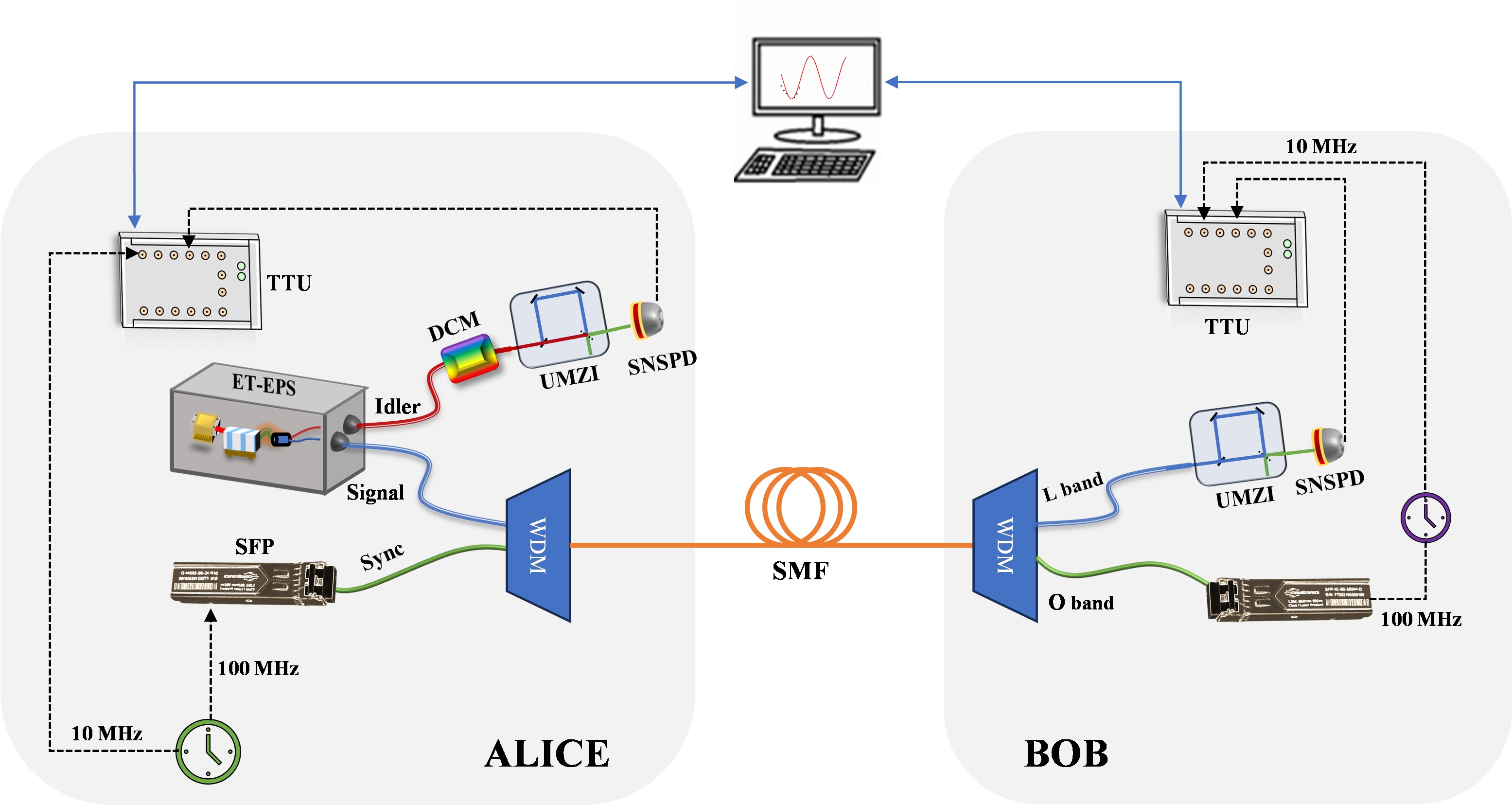}
    \caption{\label{fig:F5} Setup for non-local Franson interference with co-Propagating RoF Synchronization over 50 km optical fiber.}
\end{figure}

The experimental setup for nonlocal Franson interference with co-Propagating RoF Synchronization over a 50 km SMF link is illustrated in Fig.~\ref{fig:F5}. A self-developed entangled photon pair source (ET-EBS) was employed \cite{liu_all-fiber_2023}. Photon pairs were generated via spontaneous parametric down-conversion (SPDC) in a type-II MgO-doped periodically poled lithium niobate (MgO:PPLN, HCP) waveguide, pumped by a 780~nm continuous-wave laser. By adjusting the phase-matching temperature, the signal photon wavelength was tuned to 1575 nm (L-band), corresponding to an idler photon wavelength of 1545.3 nm \cite{Xiang_NDC_2022}.
The signal photons in Alice node were transmitted through the SMF link to the Bob node, after which they passed through an unbalanced Mach–Zehnder interferometer (UMZI, MINT Kylia) and were subsequently detected by a superconducting nanowire single-photon detector (SNSPD, Photon Technology Co., Ltd.). Meanwhile, the corresponding idler photons passed through a dispersion compensation module (DCM, DCM-CB-SN-050P1FA, Proximion AB) and another UMZI in sequence before being detected locally by a second SNSPD. The arrival times of both signal and idler photons were recorded by respective time-tagger units (TTUs, Time Tagger Ultra, Swabian Instruments). In Franson interference measurements, the two arms of the UMZI are intentionally unbalanced by approximately 500 ps, which greatly exceeds the single-photon coherence time (2 ps) to suppress any single-photon interference effects. A programmable DC power supply (not shwon in Fig.~\ref{fig:F5}) was applied to one of the UMZIs, enabling controlled variation of the optical phase difference between the two interferometers. Meanwhile, coincidence counts between the signal and idler photons are recorded to obtain the interference visibility.

To establish a synchronization between the TTUs, a 100 MHz clock signal is distributed from the local (Alice) to the remote (Bob) node. The electrical clock is transmitted by a RoF converter that uses a Small Form‑factor Pluggable (SFP) optical module (SFP‑GE‑LH60‑SM1310, 1310 nm) with an output power of 1 dBm. The clock signal carrier is then multiplexed with the quantum signal photons through a high-isolation wavelength-division multiplexer (WDM) and co‑propagates along the same optical fiber. At Bob node, the optical signal is reconverted into an electrical signal by a corresponding RoF receiver. Finally, the recovered clock is divided down to 10 MHz and delivered to the remote TTU, where it serves as a software‑defined reference clock to synchronize the timestamping process.

\section{Results and discussion}
\begin{figure}
    \includegraphics[width=8cm]{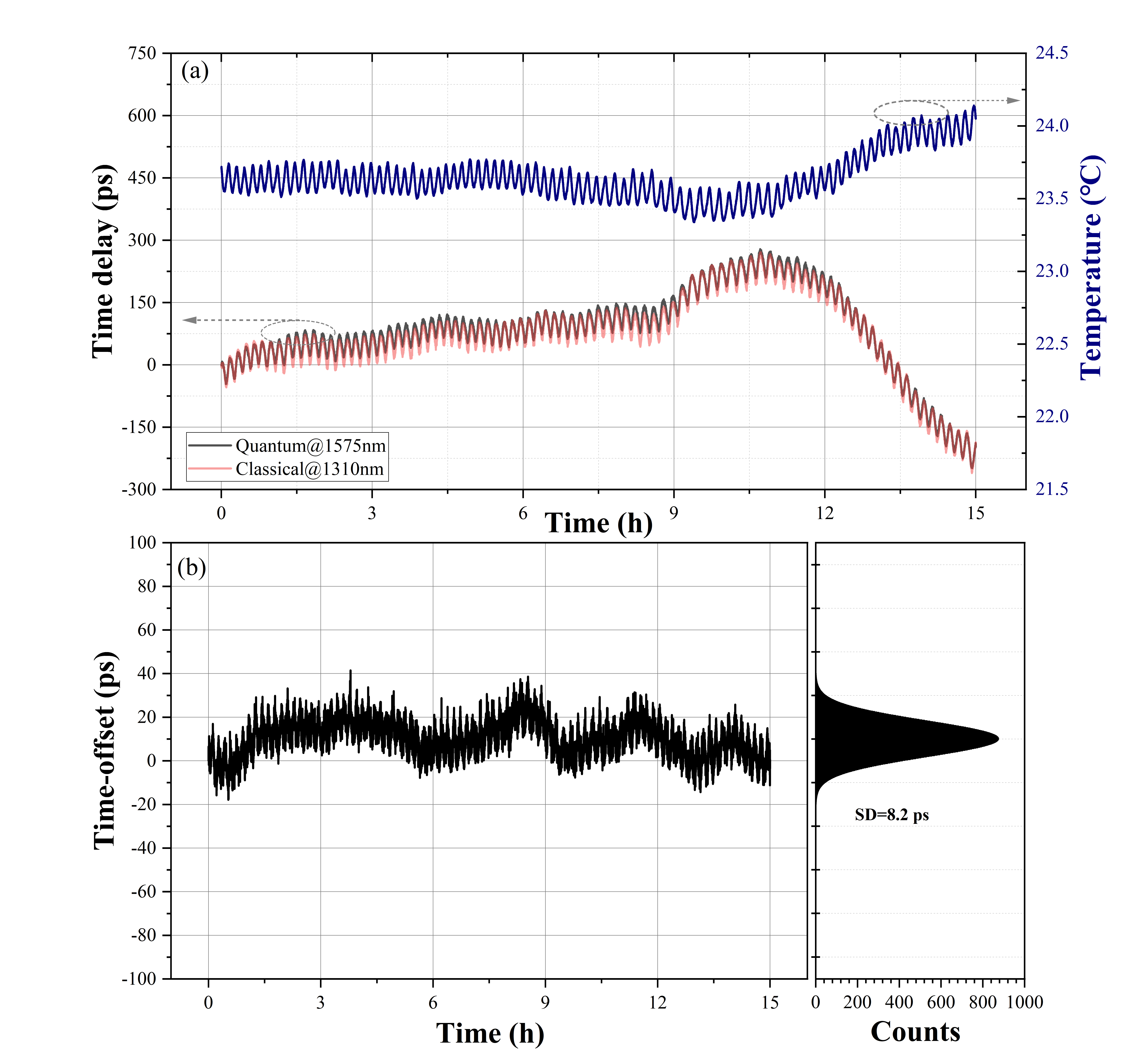}
    \caption{\label{fig:F6} (a) Relative time-delay between the 1310 nm RoF synchronization and the 1575 nm quantum signal photon over 50 km of co-propagation. (b) The differential delay with a standard deviation of 8.2 ps.}
\end{figure}

Although the classical signal and the quantum signal are transmitted through the same optical fiber, their significant wavelength difference necessitates a careful evaluation of the wavelength-dependent chromatic dispersion and temporal walk-off. Based on the setup shown in Fig.~\ref{fig:F6}, the relative delay between the RoF clock signal and the signal-idler photon pairs was measured in real-time using a single TTU. As presented in Fig.~\ref{fig:F6}(a), after 50~km fiber transmission, the delay variations of the 1310~nm synchronization signal and the 1575~nm quantum signal photon are clearly correlated. Both traces track environmental perturbations such as temperature fluctuations (navy line), as evidenced by the overlapping trends of the red and black curves, respectively. The extracted differential delay (time offset) is shown in Fig.~\ref{fig:F6}(b). It exhibits a peak-to-peak variation below 60~ps over a continuous 15-hour measurement period, demonstrating excellent common-mode noise rejection. The statistical distribution of this time offset, illustrated in the right panel of Fig.~\ref{fig:F6}(b), yields a standard deviation of 8.2~ps. 

\begin{figure}
    \includegraphics[width=9cm]{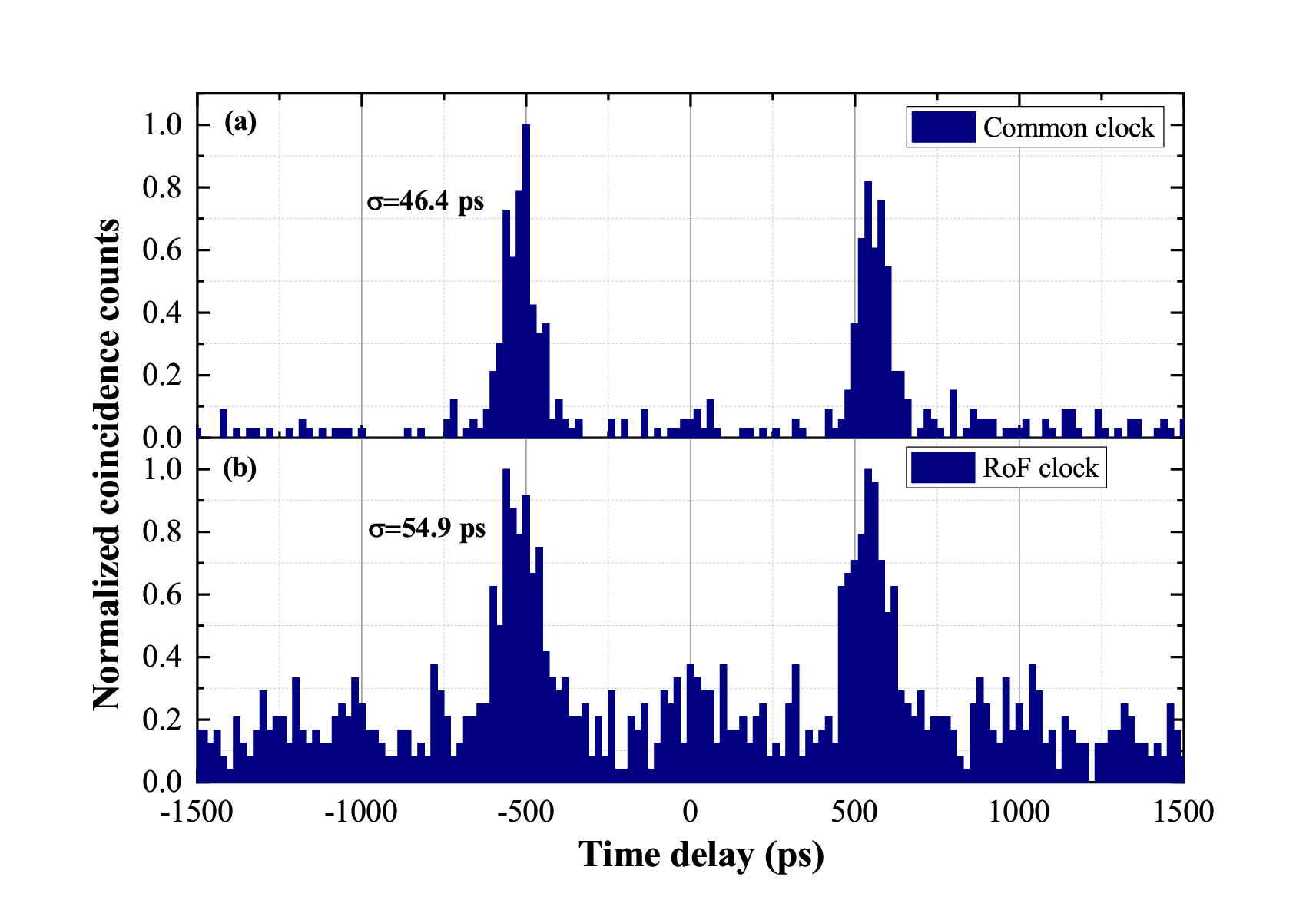}
    \caption{\label{fig:F7} Example coincidence histograms over 50 km optical fiber.}
\end{figure}
To further evaluate the synchronization performance of the co-propagating passive RoF scheme for Franson interference, we measured the coincidence counts under both common clock reference and co-propagated RoF clock conditions. As shown in the representative coincidence profiles in Fig.~\ref{fig:F7}, the left and right peaks correspond to non-interference events and exhibit stable count rates across different experimental configurations. Therefore, in the following analysis, we focus on the temporal width of the left peak. Fig.~\ref{fig:F7}(a) presents the common clock case without RoF, yielding a temporal width of 46.4 ps (standard deviation), which suggests that a small amount of uncompensated dispersion remains in the setup. In Fig.~\ref{fig:F7}(b), with RoF clock enabled, the peak  temporal width is slightly increased to 54.9 ps. The co‑propagating passive RoF scheme introduces an additional timing jitter of approximately 29.3 ps, which satisfies the stringent synchronization precision required for high-fidelity Franson interference in the theoretical section.
\begin{figure}
    \includegraphics[width=9cm]{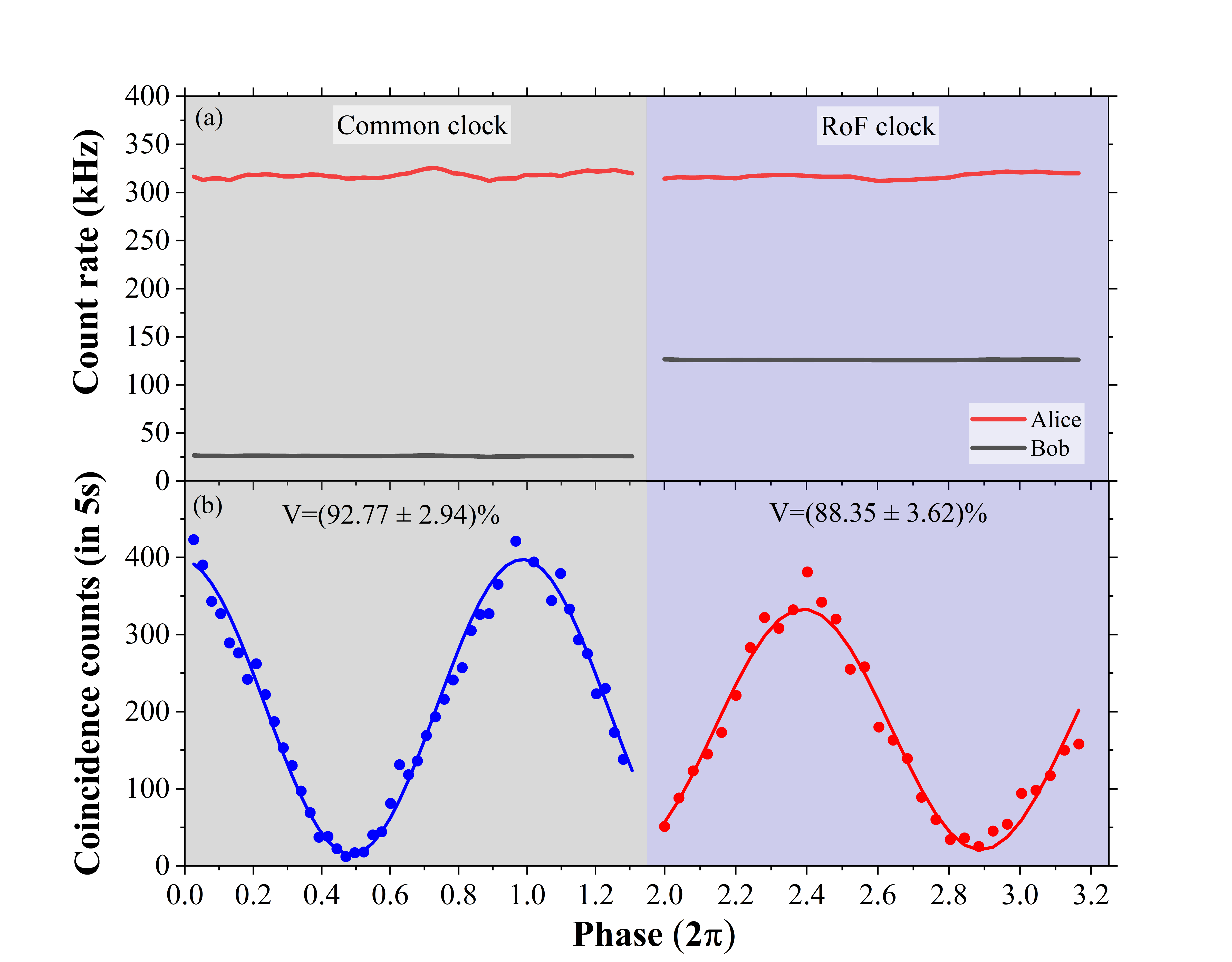}
    \caption{\label{fig:F8} Experimental results for non-local Franson interference with co-propagating RoF synchronization over 50 km of optical fiber. (a) Single‑photon count rates of Alice and Bob with the common clock and RoF clock turned ON. (b) Coincidence rate between Alice and Bob as a function of the relative interferometer phase.}
\end{figure}

For characterizing the preserved entanglement after fiber transmission co-propagating with the classical clock, we performed Franson interference measurements. The central coincidence counts were selected using a 100~ps time window, which corresponds approximately to \(2\sigma_m\), following the optimization guideline from our theoretical analysis. In the common clock scenario without RoF synchronization, the single-photon count rates of Alice and Bob remained stable, as shown in Fig.~\ref{fig:F8}(a), indicating the absence of single-photon interference. The raw Franson interference fringe is presented in Fig.~\ref{fig:F8}(b), yielding a visibility of \((92.77\pm2.94)\%\) without any background subtraction. 

When the RoF clock was enabled, SpRS-induced noise photons increased Bob’s count rate by 100~kHz, as seen in Fig.~\ref{fig:F8}(a). Consequently, the interference visibility decreased to \((88.35\pm3.62)\%\), shown in the right panel of Fig.~\ref{fig:F8}(b). From this visibility, we calculate an S-parameter of \(S = 2.50 \pm 0.10\), which violates the Clauser–Horne–Shimony–Holt (CHSH) inequality\cite{Clauser1969} (\(S \leq 2\)) by 5 standard deviations. This clear violation confirms that energy‑time entanglement is robustly preserved throughout the long‑distance distribution via the shared fiber link. The presented visibility is comparable to values reported for other fiber-based entanglement distribution systems and is sufficient for practical QKD protocols. Future work to actively stabilize the interferometer path lengths or employ narrower spectral filtering\cite{fallon_franson_2025} could further elevate the visibility toward the theoretical limit.

\section{Conclusion}
In this work, we have theoretically analyzed and experimentally validated a passive but robust synchronization scheme for long-distance energy-time entanglement distribution, relying on the co-propagation of a open-loop RoF clock signal within the same optical fiber. Our theoretical model quantified the influence of timing synchronization error on Franson interference visibility and provided a clear criterion for selecting the optimal coincidence window. Experimentally, we demonstrated that a substantial wavelength separation of 265 nm between the classical clock and the quantum channel does not prevent high-precision synchronization, achieving $\sim$30 ps timing jitter over 50 km of fiber. Using this integrated co-propagation architecture, we observed non-local Franson interference with a raw visibility of \((88.35\pm3.62)\%\), leading to a clear violation of the CHSH inequality. These findings confirm that the RoF technique offers a stable, low-cost and infrastructure-efficient synchronization method for metropolitan-scale quantum networks. This work thereby advances the practical integration of entanglement-based quantum communication protocols into existing optical communication networks, paving the way for secure quantum key distribution and distributed quantum sensing over hybrid quantum-classical fiber infrastructures.

\nocite{*}

\bibliography{ref}

\end{document}